\documentstyle[epsfig]{aipproc}

\begin{document}
\title{Recent Progress in Baryogenesis}

\author{James M.\ Cline}
\address{Department of Physics, McGill University\\
Montr\'eal, Qu\'ebec H3A 2T8 Canada}

\maketitle

\begin{abstract} I give a synopsis, specifically aimed at nonexperts, of
some of the recent developments in electroweak baryogenesis.  The focus of
the talk is on the present status of electroweak baryogenesis in
supersymmetric models, since this is a plausible and realistic possibility
that is currently being probed by experimental searches for the Higgs
boson and the top squark.  The question of whether it is viable to have a
period of color-breaking during the electroweak phase transition is also
discussed. \end{abstract}

\vskip-4in
\rightline{McGill/99-02}
\vskip4in

\section*{Baryon Asymmetry of the Universe: Review}

The baryons within the observed universe appear to consist essentially
only of matter and not antimatter.  There are no regions of gamma ray
emission which would correspond to the collision between a galaxy made of
normal matter with one made from antimatter, for instance.  Moreover, the
theory of Big Bang nucleosythesis gives predictions for the abundances of
primordial helium (both $^4$He and $^3$He), deuterium and lithium, if the
ratio of baryons to photons in the universe is in the range
\[
	2 \lesssim \eta_{10}\lesssim 6;\qquad \eta_{10}\equiv {n_B\over
	n_\gamma}\times 10^{10}\, .
\]
If we do live in a universe with equal quantities of matter and
antimatter, then the two must somehow be separated on distance scales
greater than the present Hubble length, that is, the size of the presently
observable universe.  The simplest assumption is that indeed the whole
universe has the same preponderance of matter over antimatter.  It is
a deep mystery of cosmology why the abundance of baryons should be this
peculiar number, $10^{-10}$, relative to the number of photons.

The mystery is intensified by the fact that the most natural initial
condition of the universe, at the time of the big bang or shortly
thereafter, {\it is} to have equal numbers of baryons and antibaryons,
\[
	n_B = n_{\bar B}\, ,
\]
implying that the net baryon number of the universe was zero.  There are
several reasons for believing this.

\begin{enumerate}

\item Baryon ($B$) and Lepton ($L$) conservation are {\it accidental}
symmetries of the Standard Model (SM): there is no good reason for them to
be exact.  For example, a dimension 9 operator consisting of 6
right-handed quark fields in a color singlet state,
\[
	\Lambda^{-5} (u\,d\,d)^2
\]
is allowed by the gauge symmetries, and would lead to neutron-antineutron
oscillations at some rate suppressed by the large mass scale $\Lambda$.
At sufficiently high temperatures $T\sim\Lambda$ in the early universe,
however, the effects of such a baryon number violating operator would
be unsuppressed.

\item {\it Sphalerons}, present within the SM itself, violate $B$ and $L$
(but not $B-L$).  These are the lowest energy field configurations with
Chern-Simons number $1/2$, intermediate between neighboring $N$-vacua of
the $SU(2)$ electroweak gauge theory.  (A sphaleron can be thought of as
the $t=0$ slice of an instanton, such as occur in QCD.  In QCD it is
chirality, rather than $B+L$, which is anomalously violated.)  Local
transitions which go from one $N$-vacuum to a neighboring one must pass
through a sphaleron-like configuration.  Because of the triangle anomaly
in the baryon and lepton currents, each such transition is accompanied by
9 quarks and 3 leptons.  At zero temperature, the energy barrier between
the $N$-vacua, which is the sphaleron energy, is near 10 TeV, and the
tunneling rate for the anomalous transition is so slow as to be entirely
irrelevant.  But at temperatures above that of the electroweak phase
transition, $\sim 100$ GeV, this energy barrier disappears, and sphaleron
transitions are fast compared to the Hubble expansion rate in the early
universe. 

\item Grand Unified Theories also violate B and L through heavy gauge
boson ($X$) vertices of the form $X q e$ or $X \bar q q$.  The presence
of both such interactions prevents one from assigning a conserved $B$ or
$L$ number to the $X$ boson.

\end{enumerate}

Because of these sources of $B$ violation, we would expect that any
initial $B$ or $L$ asymmetry that might have been present initially would
be quickly wiped out, giving
\[
	B = L = 0
\]
as the effective initial condition at the high temperatures of the
very early universe. Therefore something must have happened between now
and then to produce the observed baryon asymmetry.

It was realized by Sakharov in 1967 
that three things are needed to
spontaneously generate the baryon asymmetry: baryon number violating
interactions, $CP$ (particle-antiparticle symmetry) violation, and loss of
thermal equilibrium for the $B$-violating interactions.  The first two
features are present in the SM by virtue of sphalerons and the phase in
the CKM matrix for the quarks.  This source of $CP$ violation is
however too weakly coupled to the mechanism of baryon production to
produce a large enough asymmetry.  The third condition is unfulfilled in
the SM model: the transition between the symmetric and the broken
phase of the electroweak theory is so continuous that it is not a  
phase transition at all.  Sphaleron interactions, although they
eventually drop out of thermal equilibrium, do so too gradually to allow
the universe to go from a state of $B=0$ to one of nonzero $B$.

We therefore need new physics beyond the SM for baryogenesis.  
Many particle physicists believe that supersymmetry is the most natural
direction in which to look for such new physics, and I will also take that
point of view.  Moreover, I will confine my remarks to electroweak
baryogenesis, even though much interesting work on nonelectroweak
baryogenesis scenarios has been done in the last year\cite{others}.

\section*{Electroweak Baryogenesis in the MSSM}

\begin{figure}[b!] 
\centerline{\epsfig{file=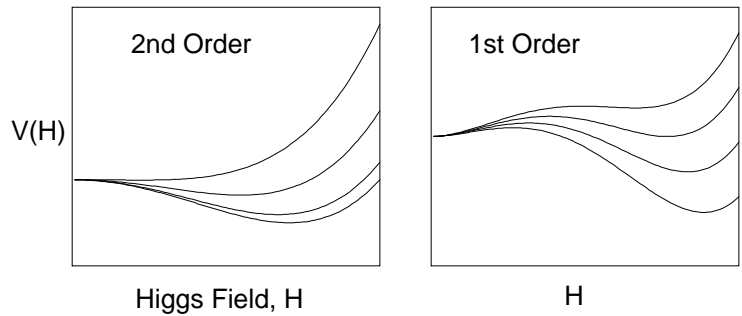,height=2in,width=4.5in}}
\vspace{10pt}
\caption{Higgs potential near a second order (left) or a first order
(right) phase transition, for a series of temperatures.}
\label{fig1}
\end{figure}

Electroweak baryogenesis within the minimal supersymmetric standard model
(MSSM) has evolved far beyond the status
of being just a rough, qualitative theory.  Rather, it has been the
subject of intense, highly quantitative scrutiny, thanks to the fact that
it is so much in the realm of currently testable physics.  The way it must
work is rather well-defined.  First, the electroweak phase transition must
be first order, which requires the Higgs field potential to have a barrier
between the symmetric phase minimum at $H=0$ and the true vacuum state
with $H\neq 0$, as shown in figure \ref{fig1}.  Under these conditions,
the phase transition proceeds by the nucleation of spherical bubbles
containing the new $H\neq 0$ phase within.  The bubbles quickly expand and 
fill the universe with the broken phase.

\begin{figure}[t!] 
\centerline{\epsfig{file=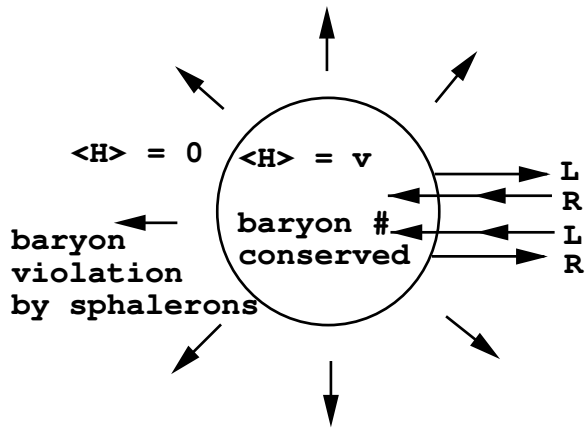,height=2.2in,width=3in}}
\vspace{10pt}
\caption{The expanding bubble wall during a first order electroweak phase
transition.}
\label{fig2}
\end{figure}

Between the initial nucleation and the completion of the phase transition,
particles in the high-$T$ plasma are encountering the expanding bubble
walls (figure \ref{fig2}).  Most particles are massless outside the bubble
and massive within, so the bubble wall behaves like a quantum mechanical
potential which can scatter the particles.  In general there is partial
reflection and transmission of particles at the wall, with left-handed
particles reflecting into right-handed ones, since spin is conserved,
and vice versa. 

It is in the reflection process that $CP$ violation is important.  If
$CP$ is violated on the wall, then particles and antiparticles can have
unequal reflection probabilities.  The same is true for left-handed and
right-handed fermions.  Thus an excess of left-handed quarks versus
antiquarks can build up in front of the wall (compensated by an equal
and opposite excess of right-handed quarks so that net baryon number is
still zero at this point).

The asymmetry between $q_l$ and $\bar q_l$ increases the free energy of
the plasma locally, and sphalerons try to minimize this energy by
destroying the $q_l$-$\bar q_l$ asymmetry to some extent, redistributing
it among the other species of quarks and leptons.  This biases the
sphalerons to preferentially create an excess of baryons or antibaryons,
which resides in front of the wall for a time, but eventually falls inside
the bubble and becomes the baryon asymmetry of the universe (BAU) that is
observed today.  However the BAU can survive inside the bubbles only if
sphaleron interactions are essentially shut off.  This is because the $CP$
asymmetry is not operative inside the bubbles (since chirality is not a
good quantum number in the broken phase) to insure that sphalerons act
preferentially to make only baryons or antibaryons.  If sphalerons are not
turned off, they will erase whatever BAU is created. 

The above idea is highly constrained.  For example, it does not
appear feasible to alter the basic mechanism by replacing bubbles with
other field configurations such as cosmic strings, despite the fact that
this possibility has been widely discussed in the literature.  We have
recently shown \cite{CEMR} that the cosmic string scenario, when
scrutinized more carefully, underproduces the baryon asymmetry by 10
orders of magnitude.  The reason is essentially that $CP$ violation at the
string walls is proportional to the string velocity squared, $v^2$, while
the density of strings in the network scales like $v^{-2}$.  It was
previously assumed that the strength of $CP$ violation and the string
density could be independently varied. 

\begin{figure}[t!] 
\centerline{\epsfig{file=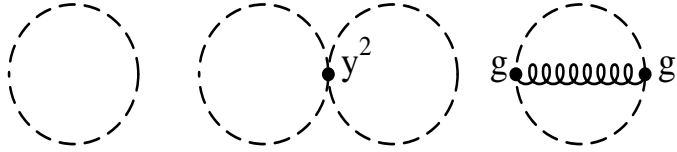,height=0.75in,width=3.5in}}
\vspace{10pt}
\caption{Some of the virtual squark diagrams contributing to $V(H)$.}
\label{fig3}
\end{figure}

However, the most important constraint on electroweak baryogenesis is the
condition for making the sphaleron interactions slow enough to preserve
the BAU, which turns out to be
\begin{equation}
\label{eq1}
	\langle H \rangle \geq T
\end{equation}
inside the bubbles, where $\langle H\rangle$ is the the field value that
minimzes the potential energy.  
Although impossible to achieve in the SM, it {\it is}
possible in the MSSM, provided that the right-handed top squark and the
lightest Higgs boson are sufficiently light.  (A light left-handed stop is
disfavored by precision electroweak considerations, namely the rho
or $T$ parameter.)  The significance of the top
squark is its large Yukawa coupling $y$ to the Higgs field,
whose potential $V(H)$ controls the phase transition.  Top squark loops 
such as those shown in figure \ref{fig3} 
contribute cubic terms of the form $-y^3 T H^3$ to $V(H)$, which make
it possible to fulfill eq.\ (\ref{eq1}).  This can be understood in terms
of the standard expression for the free energy of a relativistic gas of
bosons, represented by the one-loop diagram in figure \ref{fig3}, 
which when expanded in powers of the boson mass over temperature
contains a term proportional to $m^3 T$.  Since the mass of the squark
depends on the Higgs field, this increases the magnitude of the 
negative $H^3$ term in $V(H)$.

We have studied the EWPT in the MSSM using the two-loop effective
potential, and varying the parameters of the MSSM randomly over many
thousands of values to search for those which give a strong enough phase
transition \cite{CM}.  Our results are in reasonable agreement with
several similar studies \cite{JRE}-$\!\!$\cite{LR}, in predicting the mass
ranges for
the Higgs boson and the stop to be
\begin{eqnarray}
\label{eq2}
	85 {\rm\ GeV} &<& m_h < 107-116 {\rm\ GeV}\, ;\\
	120 {\rm\ GeV} &<& m_{\tilde t_R} < 172 {\rm\ GeV}\, .\nonumber
\end{eqnarray}
The lower bound on $m_h$ is from the latest $L3$ (LEP) experimental limit
\cite{L3} (the MSSM version of this bound is weaker than in the SM), and
the upper bound is a function of the heavy stop mass, because its
radiative corrections to $m_h$ grow logarithmically with $m_{\tilde
t}$.

One might hope that the chances of observing such a light stop at the
Tevatron would be good.  But experimental limits on $m_{\tilde t_R}$
often depend on other unknown MSSM parameters that determine which
production channels are open to $\tilde t_R$.  For example, the
greatest sensitivity to squarks is when they are produced along with
gluinos, but if the gluino mass exceeds $\sim$300 GeV then light
squarks are difficult to identify. \cite{D0}
  Similarly, the baryon production
mechanism at the bubble wall prefers values of the $\mu$ parameter and
Wino mass $m_2$ in the range $|\mu|\sim m_2\sim 100$ GeV, which would
inspire hopes for the imminent discovery of the chargino
\cite{CQRVW,CJK}.  This range of values is in fact excluded by chargino
searches for $\mu>0$, but not for $\mu < 0$.

For these reasons, the most foolproof experimental test of electroweak
baryogenesis is the search for the light Higgs boson.  The remaining
window (\ref{eq2}) is rapdily being closed by the LEP2 run at CERN.  It is
a pity, though, that LEP2 will terminate with a limit of only $m_h <
107$ GeV, leaving a small unexcluded window of Higgs mass values.  However
it is possible that a better theoretical determination of the Higgs mass,
using the renormalization group, could help to close this window. 

\section*{Could color have been briefly broken?}
One of the interesting possibilities that has been suggested in connection
with electroweak baryogenesis with a light stop is that the SU(3) gauge
group of QCD was temporarily broken.  In order to get a sufficiently light
right-handed stop, its bare mass-squared parameter must have been
{\it negative}, so that the field-dependent stop mass has the form
\[
	m^2_{\tilde t_r} = -\widetilde m^2_U + y^2 |H|^2 
	+ {g_s^2\over 6}\tilde |t_R|^2
\]
This makes the symmetric vacuum unstable toward condensation of the 
stop field in some random direction in color space.  If $-\widetilde
m^2_U$
is sufficiently negative, it can be energetically preferable to have a
period of color breaking before the normal electroweak vacuum state takes
over.  In this case one would have the sequence
\def\aone{\mathrel{\hbox{\rlap{\hbox{\lower10pt\hbox{$\,1$}}}\hbox{$\to$}}}}
\def\atwo{\mathrel{\hbox{\rlap{\hbox{\lower10pt\hbox{$\,2$}}}\hbox{$\to$}}}}

\[
	(H,\ \tilde t_R) \ = \ (0,\ 0) 
 {\,\aone\,} (0,\, v_{\tilde t}) {\,\atwo\,} 
	(v_h,\, 0)
\]
of phase transitions, which would change our view of cosmological history
in a very interesting way.  For example, the second transition tends to be
very strong, which is favorable for baryogenesis.  
But there is an energy barrier impeding this second
stage of the transition, due to a positive term in the potential
\[
	y^2 |H|^2 |\tilde t_R|^2
\]
whose presence is mandated by supersymmetry.  It could happen that the
rate of tunneling from the color-broken to the electroweak phase is so
small that it will never happen in the history of the universe. 
Guy Moore has made the following conjecture: if $m_{\tilde t_r}$ is ever
small enough for transition 1 to take place, then the universe gets
stuck in the color broken phase and never completes transition 2.  Although
preliminary studies of this question have been done, it deserves a 
more careful treatment.

We have undertaken such a study, by constructing the full two-loop
effective potential $V(H,\,\tilde t_R)$ for the Higgs and stop fields, and
computing the nucleation rate for the most likely bubbles interpolating
between the color-broken and electroweak phases \cite{CMS}.  This involves
finding the path in the $(H,\,\tilde t_R)$ field space along which the
bubble evolves, which gives the lowest bubble energy, hence the fastest
rate of transitions.  The field equations with boundary conditions
$(0,\,v_{\tilde t})$ and $(v_h,\,0)$ at the respective ends must be solved
along this path.  One needs a value of bubble energy over temperature
smaller than $E/T \cong 180$ to get a tunneling rate per unit volume,
$\sim T^4 e^{-E/T}$, that is competitive with the Hubble rate per Hubble
volume, ${\cal H}^4$.  That is, $E/T$ must be less than $4 \ln(T/{\cal
H})$.  However, we find that even when all MSSM parameters are adjusted to
the values that are optimal for tunneling, the exponent $E/T$ is too large
by an order of magnitude.  It therefore appears that one must go beyond
the MSSM in order to make color-breaking a real possibility just before the
electroweak phase transition.  

\section*{For the future}
Although the electroweak phase transition in the MSSM is now well
understood, the details of how baryons are produced at the bubble wall
are still controversial.  This is a highly complex phenomenon involving
quantum reflection or classical forces acting on the particles near the
wall, while they are simultaneously being scattered by other particles
in the plasma.  It is likely that the Quantum Boltzmann Equation will be
needed to put this part of the theory on a more rigorous footing.

\end{document}